\documentclass[epjCONF, onecolumn]{svjour}

\usepackage{epsfig}
\usepackage{graphics}
\usepackage[varg]{txfonts} % Times fonts

\session-title{Conference Title, to be filled}

\def\msun{{\rm M}_{\odot}}
\def\yr{{\rm yr}}
\def\myr{{\rm Myr}}
\def\Gyr{{\rm Gyr}}
\def\K{{\rm K}}
\def\kpc{{\rm kpc}}
\def\kms{{\rm km~s^{-1}}}

\def\lsim{\,\lower.7ex\hbox{$\;\stackrel{\textstyle<}{\sim}\;$}}

\begin{document}
%
%\title{Galactic fountains and the corona of the Milky Way}
\title{Fountain-driven gas accretion by the Milky Way}
\author{F. Marinacci\inst{1,2,3}\fnmsep\thanks{\email{federico.marinacci@h-its.org}} \and F. Fraternali\inst{1}
\and J. Binney\inst{4} \and C. Nipoti\inst{1}\and L. Ciotti\inst{1} \and P. Londrillo\inst{5} }
\institute
{Dipartimento di Astronomia, Universit\`{a} di Bologna, via Ranzani 1, 40127 Bologna, Italy
\and 
Heidelberger Institut f\"{u}r Theoretische Studien, Schloss-Wolfsbrunnenweg 35, 69118 Heidelberg, Germany  
\and 
Zentrum f\"{u}r Astronomie der Universit\"{a}t Heidelberg, Astronomisches Recheninstitut, 
M\"{o}nchhofstr. 12-14, 69120 Heidelberg, Germany
\and
Rudolf Peierls Centre for Theoretical Physics, Keble
Road, Oxford OX1 3NP, United Kingdom
\and 
INAF-Osservatorio Astronomico di Bologna, via Ranzani 1, 40127 Bologna, Italy
}

\abstract{
Accretion of fresh gas at a rate of $\sim 1~ \msun~\yr^{-1}$ is necessary in
star-forming disc galaxies, such as the Milky Way, in order to sustain their
star-formation rates. In this work we present the results of  a new hydrodynamic
simulation supporting the scenario in which the gas required for star
formation is drawn from the hot corona that surrounds the star-forming disc. In
particular, the cooling of this hot gas and its accretion on to the disc are
caused by the passage of cold galactic fountain clouds through the corona. 
} %end of abstract

\maketitle

\section{Introduction}

Each year a star-forming galaxy like the Milky Way converts $\sim 1~\msun$
of gas into stars and has done so at a fairly constant rate for nearly a
Hubble time \cite{Twarog80}. In addition, the gas content of
these galaxies has remained approximately unchanged throughout the Hubble time
\cite{Hopkins08}. Typically, the mass of gas contained in the thin disc can
sustain the process of star formation for a few $\Gyr$s only and thus, at any
given cosmic epoch, Milky Way type galaxies need  external gas to
be brought into the disc at a rate that compensates the conversion of gas into
stars \cite{Sancisi08}. A plausible reservoir of baryons available to
star-forming galaxies, which can sustain such an accretion rate for a Hubble
time, is the virial-temperature corona in which they are embedded. In this work
we present a new hydrodynamic simulation showing that the interaction between
the cold galactic fountain clouds (clouds ejected from the mid-plane of the
galaxy by supernova explosions \cite{Bregman80}) and the hot corona leads to the
cooling and the accretion of the latter on to the star-forming disc at a rate
comparable to that at which gas is transformed into stars.

\section{How coronal gas is accreted}

Our previous numerical simulations (see \cite{Marinacci10} for further details
and \cite{Marinacci11} for the dynamical consequences of the interaction)
indicate that the galactic fountain clouds effect the transfer of gas from the
hot corona to the thin disc through the following steps: (i) the
Kelvin-Helmholtz instability strips gas from fountain clouds, (ii) in their
turbulent wakes, the stripped high-metallicity gas mixes with comparable amounts
of low-metallicity coronal gas and the cooling time of the engulfed coronal gas
becomes shorter than the cloud's orbital time, (iii) knots of cold gas form and
accrete on to the disc in a dynamical time. 

\begin{figure}
\vspace{-0.2cm}
\resizebox{0.7\textwidth}{!}
{ 
 \includegraphics[viewport=40 506 575 770,clip=true]{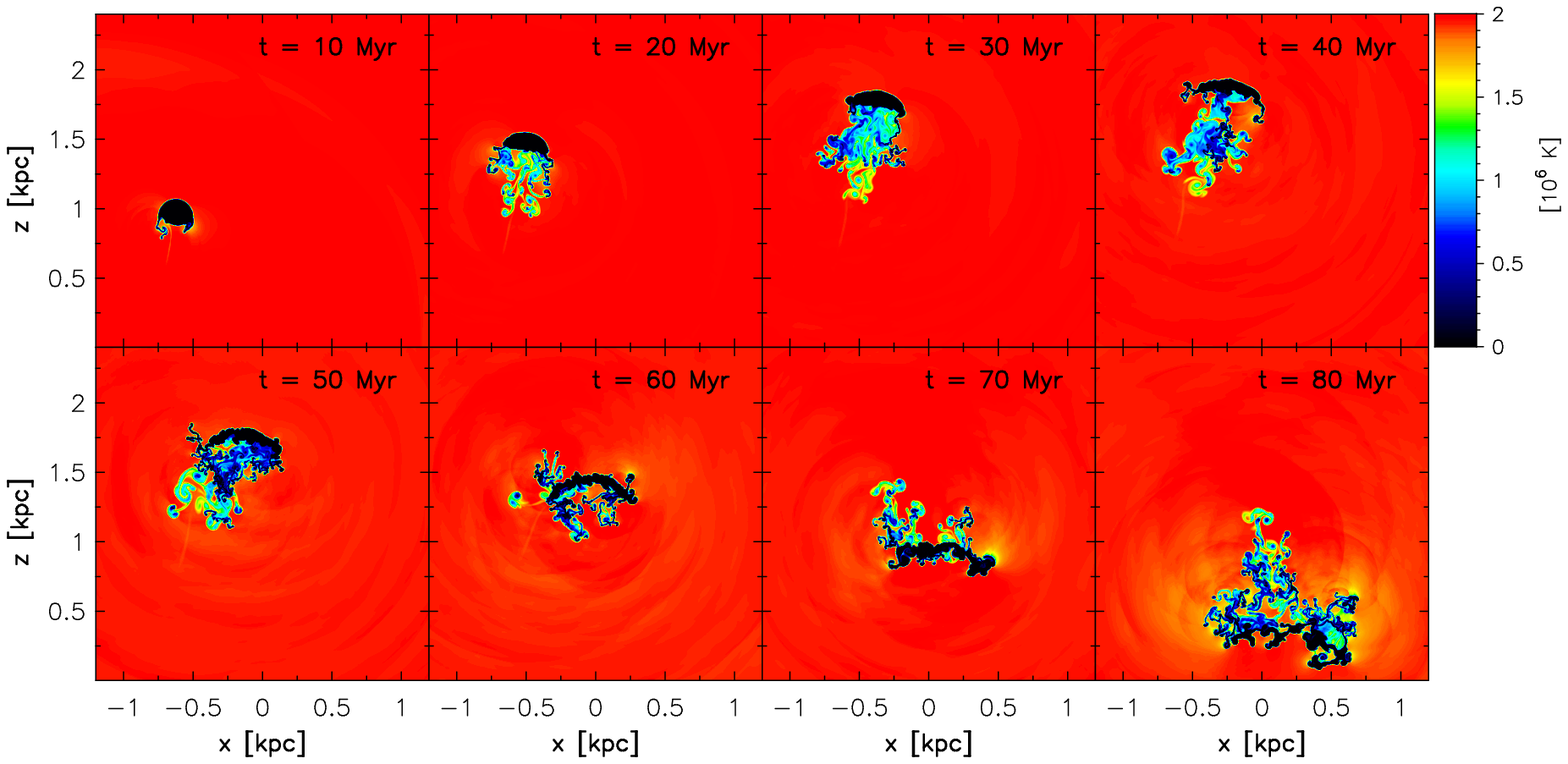}
}\hspace{-0.5cm}
\resizebox{0.33\textwidth}{!}
{
\includegraphics{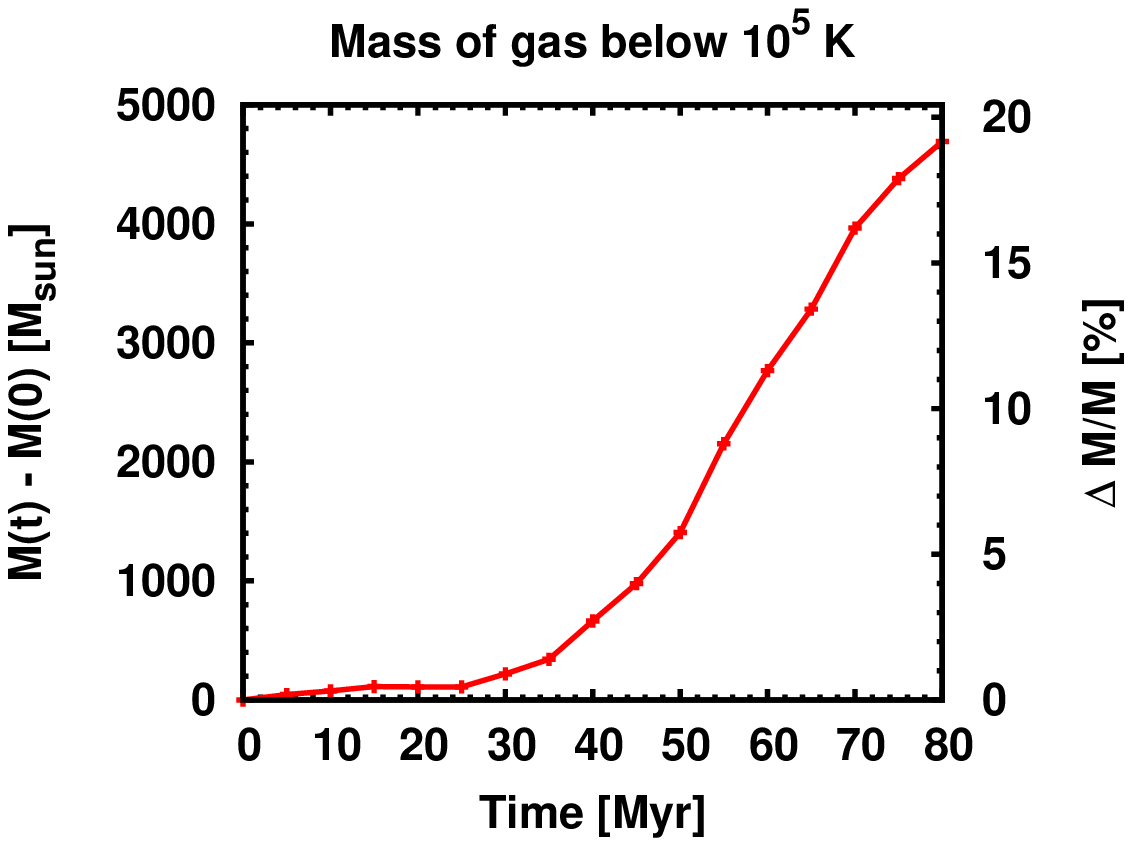}
}
\caption{\textit{Left}: Temperature snapshots for a typical fountain cloud
moving through a hot hydrostatic corona (${\rm T_{cor}} = 2\times 10^6~\K$) in
the vertical gravitational field of the Milky Way at the solar circle. Note the
formation of a turbulent wake containing knots of cold gas. \textit{Right}:
Evolution of the mass of cold (${\rm T} < 10^5~\K$) gas for the same simulation.
The mass of cold gas increases throughout the simulated time due to the
condensation of the corona.} 
\label{fig1}
\end{figure}

This situation is illustrated in Fig.~\ref{fig1} (left panel), which presents
temperature snapshots of a new 2D simulation of a typical fountain cloud moving
through a hot isothermal corona in hydrostatic equilibrium with the vertical
gravitational field of the Milky Way at the solar circle. The cloud is shot from
the bottom edge of the computational domain (where the star-forming disc is
supposedly located) with an initial velocity $v_0 = 75~\kms$, at an angle of
$10^\circ$ with respect to the vertical direction. Due to the presence of the
gravitational field, the cloud describes approximately a parabolic orbit and in
$\sim 80~\myr$ falls back to the disc. Knots of cold gas cooling from the corona
form in the cloud's wake and closely follow the motion of the main body of the
cloud. Note that the gas below $10^5~\K$ is mostly located within $\approx 1
~\kpc$ or less of the leading edge of the cloud. This suggests that the cooled
gas can be accreted on to the disc in a cloud's dynamical time to
feed the star formation. To estimate the global accretion rate on to the disc,
we computed the evolution of the mass of gas below $10^5~\K$ (Fig.~\ref{fig1},
right panel). As a consequence of the condensation of the coronal gas, this mass
increases with time throughout the simulation. When this behaviour is extrapolated
to the whole Milky Way halo, a global accretion rate of $\sim
1~\msun~\yr^{-1}$ is obtained, in agreement with our earlier results 
(see also \cite{MFB11}).

Therefore, we propose that positive feedback from supernovae, 
which drive the galactic fountain, 
is the mechanism that sustains star formation in  galaxies like the Milky
Way,
at least for redshifts $z \lsim 1$ \cite{FB08}. This
scenario also explains how star-forming galaxies can accrete gas from the hot
intergalactic medium, represented by virial-temperature coronae,
notwithstanding the thermal stability of these structures \cite{BNF09}.

\begin{acknowledgement}
We acknowledge the CINECA Award N. HP10CINJ0, 2010 for the availability
of high performance computing resources and support.
This work was supported by the MIUR grant PRIN 2008 and partially by
the Research Centre ``The Milky Way System'' (SFB 881) of the DFG.
\end{acknowledgement}

\end{document}